\documentclass[12pt,preprint2]{aastex}
\usepackage{natbib}
\usepackage{graphicx}
\usepackage{wasysym}
\usepackage{txfonts}
\def\fH2{\mbox {f$_{{\rm H}_2}$}}

\def\EBV{\mbox{\rm{E(B-V)}}}

\def\nH2{\mbox{${\rm n}(\HH$)}}
\def\enH2{\mbox{$n_{(\HH$)}}}

\def\pccc{~{\rm cm}^{-3}} 
 
\def\pcc {\mbox{${~{\rm cm}^{-2}}$}}
\def\taup {\mbox{$\tau_{353}$}}

\def\Tsub#1 {\mbox{${\rm T}_{\rm #1}$}}
\def\TK  {\Tsub K }

\def\p{\mbox{$^+$}}

\def\h13cop{\mbox{{H$^{13}$CO\p}}}

\def\c3h2{\mbox{C$_3$H$_2$}}

 \def\R0{R$_0$}
\def\G0{\mbox{G$_0$}}

\def\ddeg{{}^\circ\kern-.1em}

\def\kms{\mbox{km\,s$^{-1}$}}

\def\m1{\mbox{$^{-1}$}}

\def\E#1 {$10^{#1}$}
\def\E#1 {E{#1}}
\def\P#1,{$\nH2\TK~=~#1\times~10^4\pccc$~K}
\def\ec#1,#2,#3,{#1\,(#2)\E{#3}}

\def\H3{\mbox{H$_3$}}

\def\RH2{\mbox{R$_{\rm G}$}}
\def\g13{\mbox{g$_{13}$}}

\def\kHeH2{\mbox{$k_{ He-\HH}$}}

\def\tim#1,#2{\mbox{{$#1\times10^{#2}$}}}

\def\WHI{\mbox{$\Upsilon_{{\rm HI}}$}}
\def\ESFD{\EBV$_{\rm SFD}$}
\def\EQSO{\EBV$_{\rm QSO}$}

\newcommand{\emm}[1]{\ensuremath{#1}}   
\newcommand{\emr}[1]{\emm{\mathrm{#1}}} 

\newcommand{\HI}{\emr{HI}} 
\newcommand{\HH}{\emr{H_2}}

\slugcomment{generated \today}

\shorttitle{Ebv+HI=}
\shortauthors{Harvey Liszt}

\begin{document}


\title{Optical Reddening, Integrated HI Optical Depth, Total Hydrogen Column Density}


\author{Harvey Liszt}
\affil{National Radio Astronomy Observatory \\
        520 Edgemont Road, Charlottesville, VA 22903-2475}

\email{hliszt@nrao.edu}





\begin{abstract}
Despite the vastly different angular scales on which they are measured,
the integrated $\lambda$21 cm \HI\ optical depth measured interferometrically, 
\WHI, is a good proxy for the optical reddening derived from IR dust emission, with 
\WHI\ $\propto$ \EBV$^{1.10}$ for 0.04 mag $\la$ \EBV\ $\la$ 4 mag.  
For \EBV\ $\la 0.04$ mag or \WHI\ $< 0.7$ \kms, less-absorbent 
warm and ionized gases assert 
themselves and $\tau$(HI) is a less reliable tracer of \EBV.  
The \WHI-\EBV\ relationship can be inverted to give a broken power-law 
relationship between the total hydrogen column density N(H) and \WHI\ such that 
knowledge of \WHI\ alone predicts N(H) with an accuracy of a factor 1.5 
($\pm 0.18$ dex) across two orders of magnitude variation of \WHI.  
The \WHI-N(H) relation is invariant under a linear rescaling of the reddening 
measure used in the analysis  and does not depend on knowing properties
of the HI such as the spin temperature. 
\end{abstract}



\section{Introduction}

The presence of a pervasive gaseous interstellar medium (ISM) was demonstrated 
over the course of several decades after \cite{Har04} noted the presence
of narrow, stationary absorption lines of NaI and CaII in the spectrum 
of the binary star $\delta$ Orionis.  Ruling out the possibility that the lines 
were circumstellar took time and considerable ingenuity.  \cite{Pla22} showed that 
the stationary lines observed toward HD47129 were at rest with respect to the 
Local Standard of Rest (LSR), rather than the star.  \cite{Stru28} noted
the implied lumpy nature of the absorbing gas and showed that the integrated 
strength of the NaI D lines increased with stellar distance. \cite{PlaPea30}
showed that the Oort A-constant for the galactic rotation of the interstellar 
gas was just half that of the stars and it was generally accepted that
the absorbing medium  was distributed across the various sightlines 
and interstellar space.

The mean free path  {\it l} 
 or spatial frequency (1/{\it l}) of absorption 
features could be derived given the known stellar distances, and the discrete 
nature of the absorption lines gave rise to the notion of interstellar clouds 
of uncertain size but having a 
small volume filling factor\footnote{For uniform spheres 
the volume filling factor can be expressed as the ratio of diameter D to 
mean free path, f$_{\rm vol}$ = 2/3 D/{\it l}.}  
The discrete nature of the intervening matter was greatly reinforced by M\"unch's 
analyses of the statistics of stellar color excesses and absorption
line kinematics 
\citep{Mun52,Mun57}, and the persistence of interstellar clouds led \cite{Spi56} 
to infer that they were bounded by an unseen but pervasive hot medium,
the galactic corona.  However, optical absorption lines proved to be 
problematic for understanding the ISM because of the difficulties in relating 
the strengths of lines from heavily depleted trace species in uncertain
ionization states to hydrogen column densities \citep{Edd34,Edd37,SpiRou52}.

The discovery of $\lambda21$cm HI emission and the subsequent detection of 
HI absorption \citep{HagLil+55,ClaRad+62,Cla65} and self-absorption \citep{Hee55} 
provided gas kinetic temperatures \citep{Fie59b} and hydrogen column densities
of HI clouds as the optically-defined diffuse clouds came to be known to radio
astronomers.  In a smart synthesis, \cite{Spi68,Spi78} put M\"unch's color 
excesses on the modern scale of optical reddening \EBV\ and related \EBV\ to 
$\lambda$21cm HI hydrogen column density, directly translating M\"unch's 
statistics to express the HI cloud column density distribution in terms of 
standard and large HI clouds having smaller and larger sizes, column densities
and mean free paths.  

Although a mean interstellar gas density of approximately 1 H-atom $\pccc$ has 
been inferred since the early days of ISM studies \citep{Oort32,Edd34}, Spitzer's 
version of M\"unch's mean color excess per unit distance, 
$<$\EBV/{\it l}$>$ = 0.61 mag/kpc, is the modern origin of the widely-quoted 
mean ISM density, 1.2 H-nuclei $\pccc$. The required conversion from reddening 
to hydrogen column density was for many years taken as 
$5.8 \times 10^{21}\pcc$ H-nuclei (mag)$^{-1}$
\citep{Spi68,Spi78,BohSav+78,SavDra+77} but we found 
N(H) = $8.3 \times 10^{21}\pcc$ \EBV\ \citep{Lis14yEBV,Lis14xEBV} and
other recent studies find ratios 
$7.7-9.4 \times 10^{21}\pcc$ H-nuclei (mag)$^{-1}$ 
\citep{HenDra17,LenHen+17,LiTan+18}. This is in part due to the suggested 
14\%\ downward scaling \citep{SchFin11} of the all-sky reddening maps 
of \cite{SchFin+98}.  

In the course of our work defining the \EBV-N(H) relation 
\citep{Lis14yEBV,Lis14xEBV}, it was required to estimate the degree of 
saturation of \HI\ emission profiles and this was accomplished by relating 
the integrated $\lambda$21 cm \HI\ optical depth 
\WHI\ $\equiv \int \tau({\rm \HI})~{\rm dv}$ (units of \kms) to 
the all-sky map of FIR dust emission-derived reddening equivalent of 
\cite{SchFin+98}.  Comparison of the 
separate relationships between the integrated HI emission and absorption with 
\EBV\ defines a mean spin temperature-\EBV\ relationship that directly indicates 
the degree of saturation in comparison with the observed HI brightness.  We 
found that optical depth effects were not responsible for the larger \EBV/N(HI) 
ratios that are seen at \EBV\ $\ga$ 0.07 mag but see \cite{FukTor+15}
for an opposing view.

Perhaps surprisingly given the vastly different angular scales on which they
were measured, \WHI\ and \EBV\ are tightly coupled.  From a combined 
sample of some 100 inteferometric measurements of \WHI\ spanning 35 years we 
showed that there was a strong and very-nearly linear power-law relationship 
\WHI\ = 14.02 \EBV$^{1.07}$ \kms\ for  0.02 $\le$ \EBV $\la$ 4 mag: the 
lower limit on \EBV\ arises because weakly-absorbing warm or ionized hydrogen 
may dominate at N(H) $\la 1-2\times10^{20}\pcc$ \citep{RoyKan+13}. 

Recognizing that there are cases where the \HI\ optical depth is measurable and 
the total hydrogen column density N(H) is not (for instance when CO 
observations are not available to estimate N(\HH)), the tight relationship 
between \WHI\ and \EBV\ suggests that
\WHI\ can be a useful proxy for N(H) if that is simply related to \EBV.  
Since our earlier discussion the SPONGE \HI\ absorption sample has become
available, adding some 50 new measurements of \WHI\ \citep{SPONGE18}, and
\cite{RoyFra+17} published a smaller sample of \HI\ absorption measurements at 
especially low \EBV.  The larger ensemble of measurements of \WHI\ is employed
in this work to better specify the \WHI-\EBV\ relationship, to gain insight
into the structure of the ISM, and to derive an empirical relationship 
that predicts N(H) from \WHI\ to an accuracy of $\pm$0.18 dex over a range of 
reddening and hydrogen column density spanning more than two orders of magnitude.
The \WHI-N(H) relation is invariant under a linear rescaling of the reddening 
measure used in the analysis. It can be used to infer other properties
of HI in the ISM, but does not depend on knowing them. 

The organization of this work is as follows.  Section 2 is a discussion of 
\WHI, \EBV\ and the total column density of hydrogen nuclei N(H), where we derive 
relationships between \WHI\ and \EBV\ and \WHI\ and N(H) and discuss their 
implications for the constitution of the diffuse interstellar medium 
(ISM).  Section 3 is a summary and discussion.  In  Appendix A we discuss
an apparent disparity between integrated HI optical depths derived from
interferometric and singledish emission-absorption measurements.

\begin{figure*}
\includegraphics[height=6.9cm]{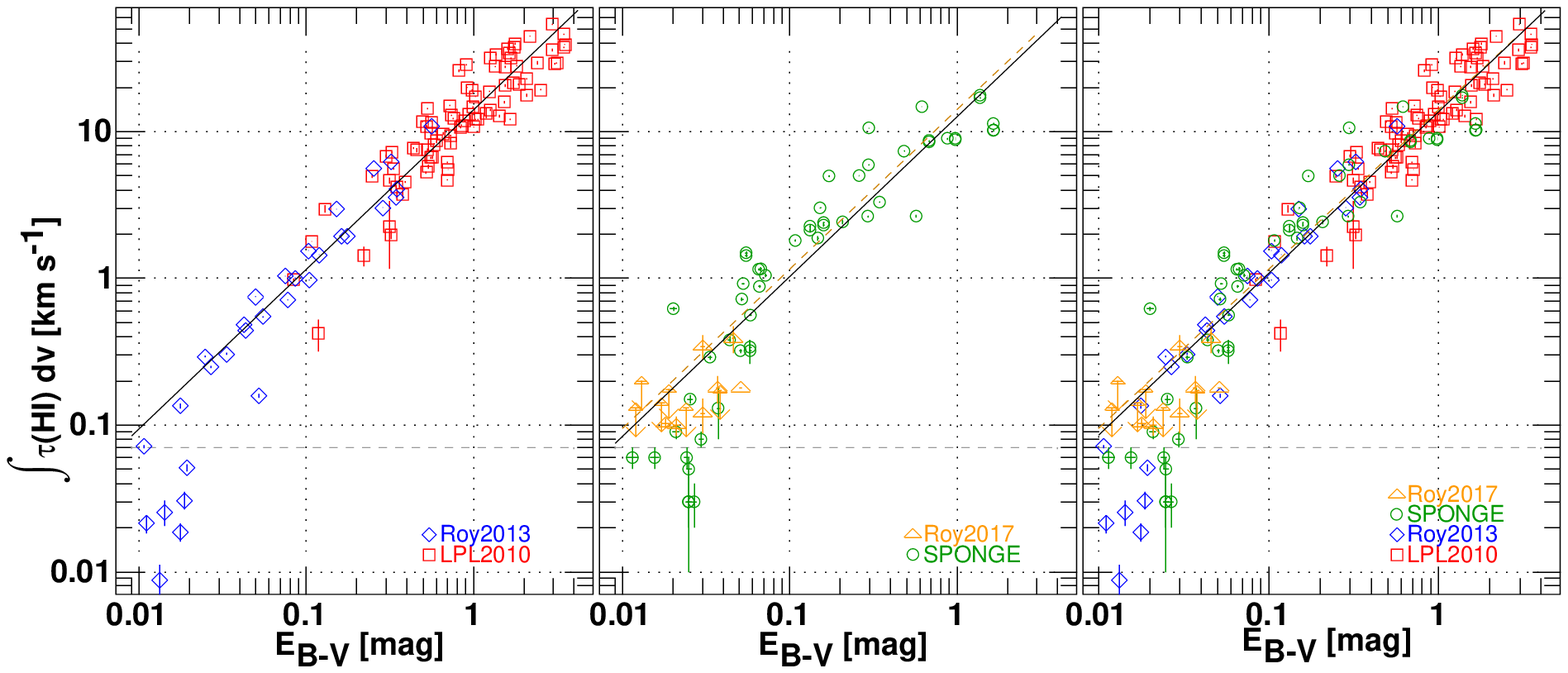}
 \caption[] {Integrated $\lambda$21 cm \HI\ optical depth for several samples,
plotted against IR dust emission-derived optical reddening from \cite{SchFin+98}.  
Left:  
Interferometric \HI\ optical depths from \cite{RoyKan+13} and the sample assembled by 
\cite{LisPet+10}, most of which is from the survey of \cite{DicKul+83}.  
Center: From the interferometric SPONGE survey \citep{SPONGE18} with data taken
at the JVLA.  Right: The merged sample from the two other panels. Solid lines 
are the power-law regression fits to the samples in each panel derived using 
data with \WHI\ $\ge$ 0.07 \kms\ and the dashed orange line in the center and 
right panels is the power-law fit derived from 107 sightlines in the panel at 
left. The regression fit for 160 sightlines with detections of HI absorption
in the merged sample is \WHI\ = 13.80\EBV$^{1.10}$ \kms. }
\end{figure*}

\section{\WHI\ and reddening}

In this work, as before, we take \EBV\ from the FIR 
dust emission-derived equivalent optical reddening of \cite{SchFin+98} (SFD), 
also occasionally denoted as \ESFD\ when such a distinction is needed.  
As we discuss, the basic result of the present work, a relation between 
hydrogen column density and integrated $\lambda$21cm HI optical depth 
is independent of a linear rescaling of \ESFD.

Figure 1 shows the relationship between \ESFD\ and various 
 interferometrically-measured samples of the integrated \HI\ optical depth
\WHI.  The left-most panel repeats the sample shown in Figure 3 of 
\cite{Lis14xEBV} using \HI\ optical depths from \cite{RoyKan+13} and the 
sample assembled by \cite{LisPet+10}, most of which is from the work of 
\cite{DicKul+83}.  The center panel shows more recent results from the new 
SPONGE survey \citep{SPONGE18} with data taken at the JVLA, and data from
the sample of \cite{RoyFra+17} at low \EBV\ observed at the JVLA and 
GMRT.  From the latter work we plot
as $3\sigma$ upper limits all results with lower statistical signficance.
The SPONGE sample has more curvature but lies 
very nearly along the regression line defined by the larger sample of 
older measurements.  

The solid line in each panel is the power-law  regression fit to the data with 
\WHI\ $\ge 0.07$ \kms\ and the dashed line in the center and right panels is 
the power-law fit derived from the 107 sightlines used in the regression fit 
at left.   The combined sample of 160 sightlines at right 
has the regression line 


$$ \WHI\ =
  (13.8\pm0.7)~\EBV^{(1.10\pm0.03)}~\kms \eqno(1) $$.

This has the very nearly the same slope and a 2\% smaller multiplier than 
we determined previously \citep{Lis14xEBV} from the sample shown at left 
using the data at \EBV\ $\ge 0.02$ mag.

As originally noted by \cite{RoyKan+13}, \WHI\ is a less reliable indicator of 
N(\HI) for N(\HI) $\la 2\times10^{20}\pcc$ (equivalently, \EBV\ $ \la 0.02-0.03$ mag) 
where the hydrogen is more likely to be warm or ionized and weakly-absorbing. 
This is also recognizable as the limiting column density defining the sample 
of damped Lyman-alpha (DLA) systems in the cosmological context of the 
Lyman-alpha forest, but galactic \HI\ remains mostly neutral even at 
\EBV\ $= 0.01$ mag as shown in Figure 1 of \cite{Lis14yEBV}.  In any case,
the full sample shows that the ability of \WHI\ to trace \EBV\ deteriorates 
at \EBV\ $\la$ 0.04 mag. 
 
\begin{figure}
\includegraphics[height=7.7cm]{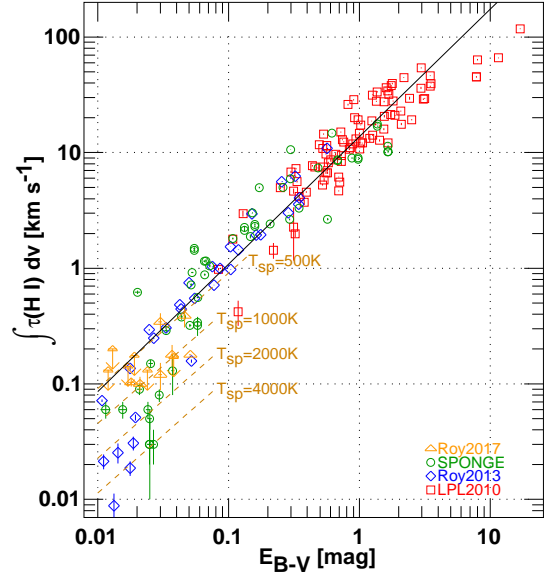}
 \caption[] {An extended view of the merged interferometric HI absorption
sample. The regression line is the same as that shown in the rightmost
panel of Figure 1, determined for $ \int\tau(\rm {\HI}) ~{\rm dv} \ge 0.07$ 
and \EBV\ $\le 4$.  Also plotted are curves of optical depth 
for N(\HI) $= 8.3 \times 10^{21} \pcc $\EBV\ and \HI\ spin temperature 
500 K, 1000 K, 2000 K and 4000 K in a uniform, isothermal gas.}
\end{figure}

\begin{figure}
\includegraphics[height=7.7cm]{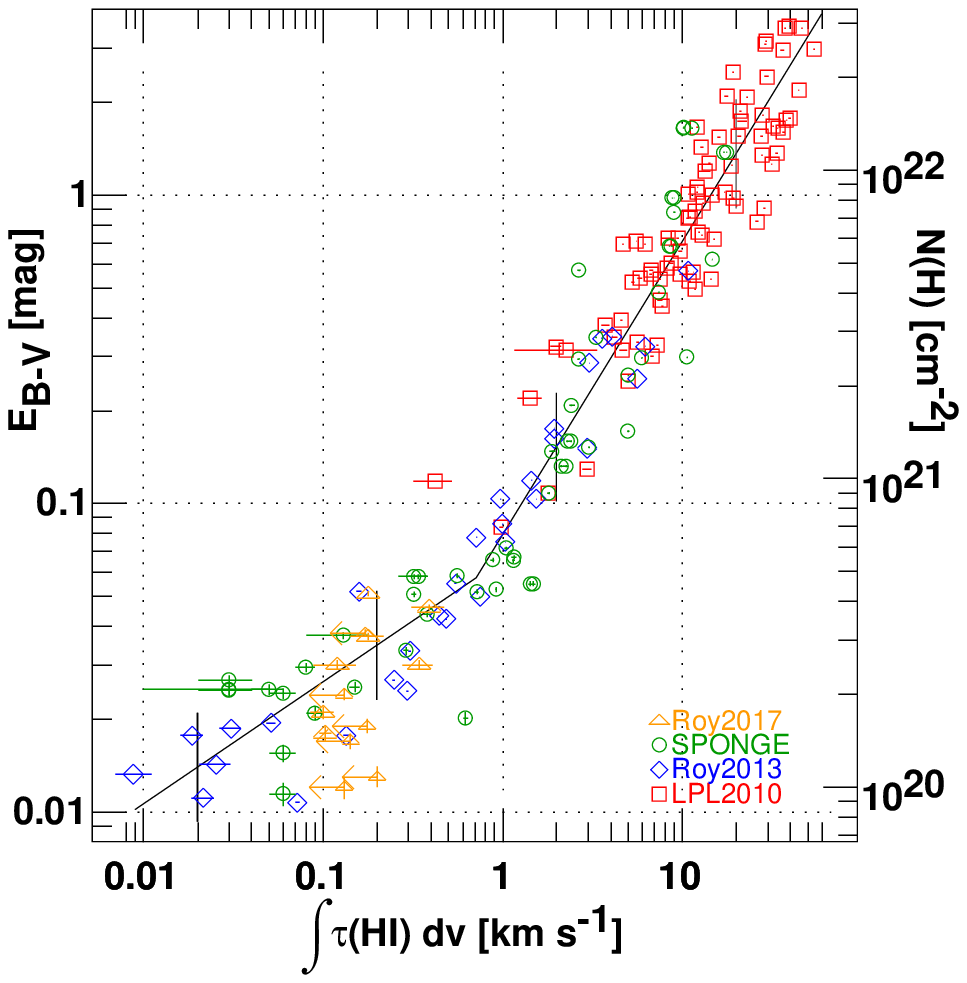}
 \caption[] {Reddening and implied N(H) vs \WHI.  The data plotted at right in 
Figure 1 are shown with the variables interchanged and fit with two power laws
meeting at \WHI = 0.72 \kms.  Shown along the fitted curves
are vertical markers denoting factor 1.5 ($\pm 1.76$ dex) variations.
The scaling to N(H) uses N(H)/\EBV\ = $8.3 \times 10^{21}\pcc$ mag$^{-1}$
and is invariant to a linear re-scaling of \EBV\ as discussed in Section 2.3.}
\end{figure}

The demarcation of the warm $\rightarrow$ cool \HI\ transition seemed more 
clearly defined and at smaller \EBV\ $\la 0.02$ mag in the older sample shown in 
the left-most panel of Figure 1. In the more recent data, 
hydrogen along sightlines with E(B-V) as large as 0.04 mag is not always dominated 
by cool, strongly-absorbing gas.  Given that sightlines with \EBV\ $\ga 0.07$ mag 
studied locally begin to show appreciable amounts of \HH\ \citep{SavDra+77}, there 
is only a narrow range of reddening over which the gas  as a whole can 
be considered to be predominantly neutral, cool, and atomic.  In this context it is also
noteworthy how weak the integrated \HI\ absorption is, as we now discuss. 

\subsection{Composition of the absorbing \HI\ and its spin temperature}


Shown in Figure 2 is an extended version of the rightmost panel of Figure 1, with 
the range in reddening extended to sightlines at very low galactic
latitude and large reddening where \ESFD\ is unreliable and the data do not 
follow the same power-law as for \ESFD\ $\la$ 4 mag.  The solid line in Figure 2
is the same regression line that is shown at right in Figure 1.
Also superposed in Figure 2 are dashed lines of constant \HI\ spin temperature 
T$_{\rm sp}$ that reproduce the plotted values of \WHI\ in a uniform atomic gas 
when N(\HI) = N(H) = $8.3 \times 10^{21}\pcc$ \EBV. 


The spin temperatures that reproduce the integrated optical depths in
a uniform gas are large compared to the emission line brightness, 
implying low optical depth. However, the spin temperatures that reproduce 
the regression line at \EBV\ $\la$ 0.1 mag are near 500 K and are unphysically 
large, given that specific determinations of the spin temperature in \HI\ 
absorption profiles consistently put the typical kinetic temperatures of individual 
kinematic components below 100 K
\citep{Cla65,DicTer+78,DicTer+79,PayTer+82,PaySal+83,HeiTro03,SPONGE18}.

This disparity between the spin temperatures that reproduce the observed integrated
\HI\ optical depths  and the kinetic temperatures of CNM imply that only a small fraction 
of the gas implied by the \EBV\ - N(H) conversion produces measurable \HI\ absorption.  
As an example, if the gas is artificially 
partitioned into an absorbing component with \HI\ spin temperature 80 K and another 
that does not absorb at all, the proportion of absorbing gas at 80 K varies from 
17\% to 23\% along the regression line. The SPONGE survey \citep{SPONGE18} 
concluded that 50\% of the \HI\ was detected in absorption. 
Overall, estimates of the molecular fraction of the diffuse ISM are in the range 
25\% - 40\% \citep{BohSav+78,LisLuc02,LisPet+10}, or roughly one-third.  If one-third
of the hydrogen in the diffuse ISM is in \HH\ and 50\% of the \HI\ is detectable 
in \HI\ absorption, the overall ionized gas fraction would be one-sixth.

\subsection{Variation of \EBV\ with N(H)}

A near-linear \WHI-\EBV\ proportionality could arise in a sufficiently well-mixed medium 
at large reddening, or if the gas observed at larger reddening were an accumulation 
of more parcels of the same gas that is seen at small reddening.  However, the mix of 
strongly and weakly $\lambda$21cm-absorbing \HI\ and unabsorbent ionized or molecular 
hydrogen is  likely to vary  over the wide range of reddening sampled in this work.  
Lines of sight with higher total \EBV\ are more likely to have a contribution from gas
parcels that locally have relatively high \EBV. 

\WHI/\EBV\ increases with \EBV\ along the slightly-superlinear regression lines
in Figures 1 and 2.  Expressing 

$$\WHI/\EBV =
  \frac{\WHI}{\rm N(H)} \times \frac{\rm N(H)}{\EBV} \eqno(2) $$

one sees that gas with higher \WHI/N(H) and/or N(H)/\EBV\ must increasingly be 
seen along the sightlines having higher total \EBV. However, \cite{PlaXI} found that 
N(H)/\taup\ decreased at higher column density, where \taup\ is the 353 GHz dust optical 
depth that can be directly scaled to \EBV\ (see Appendix A).  N(H)/\EBV\ would 
presumably change in the same way and this acts in the wrong sense, depressing 
\WHI/\EBV.   To restore the observed behaviour \WHI/N(H) would have to increase 
by an even larger amount.  This interplay between the properties of the dust and 
HI absorbing gas remain to be explored but the \WHI-\EBV\ relationship seems a 
general characteristic that should be explained by any model of the ISM.

\subsection{Conversion from \WHI\ to N(H)}

Figure 3 shows the data from the combined sample with the variables interchanged
and N(H) calculated as N(H)/\EBV\ = $8.3\times 10^{21}\pcc$ mag$^{-1}$.
The detections are well represented by two power laws, 
\EBV\ = 0.0655\WHI $^{0.395}$ mag for \WHI\ $\le$ 0.715 \kms\
and \EBV\ = 0.0788\WHI$^{0.950}$ mag for 0.715 \kms\ $\le$ \WHI\ $\le$ 60 \kms.
The break occurs at the same \WHI\ that was used as the lower-limit selection 
criterion to fit the \EBV-\WHI\ relationship of Eq. 1 as discussed in Section 2
and shown in Figure 1. The upper limits taken from the work of \cite{RoyFra+17}
do not present constraints when the data are approximated in this way.  

Shown
along the power-law fits in Figure 3 are vertical bars indicating a span of
$\pm0.15$ dex (= 2$^{1/2}$) and only a few points lie outside this range. The
expected accuracy is therefore of this order.  A Monte Carlo analysis with
gaussian random errors in \WHI\ yields a 95\% confidence limit of 0.18 dex, or 
a factor 1.5.

In terms of the column density
$$ N({\rm H}) = A \WHI^{0.395}, \WHI\ \le 0.715~\kms \eqno(3a)$$
and 
$$ N({\rm H}) = B \WHI^{0.950}, 0.715~\kms \le \WHI \eqno(3b) $$
where A $ =5.43\times10^{20}\pcc$ and B $ = 6.54\times10^{20}\pcc$.  

It is important to note that the column density-integrated HI opacity relationship 
derived in this way uses the ratio N(H)/\EBV\ = $8.3\times 10^{21}\pcc$ mag$^{-1}$
that  we determined previously, but is independent of a linear rescaling of the 
\ESFD\ values we used.  The same result would have been derived now if we had 
earlier correlated N(HI) with 
\EBV$^\prime_{\rm SFD} = 0.86$ \ESFD\ suggested by \cite{SchFin11}, 
finding N(H)/\EBV\ = $9.65\times 10^{21}\pcc$ mag$^{-1}$, and 
correlated \WHI\ with \EBV$^\prime_{\rm SFD}$ now.  Thus the ratio 
N(H)/\EBV\ = $8.3\times 10^{21}\pcc$ mag$^{-1}$ is an 
important fiducial for the N(H)-\WHI\ relationship, but the question
of whether \ESFD\ or  \EBV$^\prime_{\rm SFD}$ is the better measure of 
reddening is academic.

\section{Summary and discussion}

In Section 2 (see Figure 1) we discussed how the integrated $\lambda$21cm optical 
depth \WHI\ measured interferometrically is a good proxy for the reddening \EBV\ 
when \EBV\ $\ga 0.04$ mag or N(H) $\ga 2-3\times 10^{20}\pcc$.  Less absorbent 
ionized and warm neutral hydrogen make a noticeably larger proportional contribution 
to N(H) at \EBV\ $\la 0.04$ mag. From an updated HI absorption sample including sightlines 
from the recent SPONGE survey \citep{SPONGE18} and measurements from \cite{RoyFra+17} 
we derived the regression line \WHI\ = $(13.8\pm0.7)$\EBV$^{(1.10\pm0.03)}$ \kms.
As in our earlier work, the reddening measure we used was the IR dust-emission
derived optical reddening equivalent of \cite{SchFin+98} that we denoted as \ESFD\ 
as necessary in the text.


Comparison of the observed integrated \HI\ optical depths with the total 
hydrogen column densities implied by the ratio N(H)/\EBV\ = 
$8.3 \times 10^{21}\pcc$ mag$^{-1}$ from \cite{Lis14yEBV} in Section 2.1
showed that only a small fraction of the diffuse gas in the ISM 
is represented in \HI\ absorption, as illustrated by the artificially 
high mean \HI\ spin temperatures shown in Figure 2 for hypothetical gases 
comprised only of neutral atomic hydrogen.  These high spin temperatures can 
be reconciled with measured kinetic temperatures below 100 K if only a fraction 
of the gas absorbs strongly in HI.
Overall, the H-nuclei in the diffuse ISM are (approximately) one-third in \HH\ and
one-sixth in H\p. One-half of the gas is in neutral hydrogen atoms and  
half of these are detected in absorption according to \cite{SPONGE18}.  The fraction
of the ISM gas overall that resides in the strongly-absorbing HI cold neutral medium 
(CNM) is about 20\%. 

In Section 2.2 we analyzed the slight super-linearity of the 
\WHI\ $\propto$ \EBV$^{1.1}$ power-law relationship seen in Figure 1, expressing 
\WHI/\EBV = (\WHI/N(H)) $\times$ (N(H)/\EBV) in Eq. 2 as the product 
of separate factors characterizing the HI-related and dust-related properties 
of the gas.  At least one of those factors must increase with total \EBV\ to 
provide 
the observed 1.1 power-law slope. It is expected to find more high-column 
density parcels of colder \HI\  at higher \EBV, increasing \WHI/N(H). However, 
\citep{PlaXI} found that
N(H)/\taup\ $\propto$ N(H)/\EBV\ decreased for gas with higher N(H),
implying the need for \WHI/N(H) to increase even more rapidly with \EBV.  
The  larger issue is the extent to which N(H)/\EBV\ varies and the extent to 
which such variations contribute to uncertainty in the generally-accepted 
single value that is in use.

In Section 2.3 (see Eqs. 3a and 3b) we assumed a value of N(H)/\EBV\ =
$8.3 \times 10^{21}\pcc$ mag$^{-1}$ and inverted the  \EBV-\WHI\ 
relationship to derive a broken power-law relation between \WHI\ and N(H) 
that reliably predicts the inferred N(H) from detections of \WHI\ to 
within a factor 1.5 ($\pm$0.18 dex) for 0.01 $\le$ \WHI\ $\le$ 60 \kms. 
As noted in Section 2.3 the N(H)-\WHI\ relationship so derived is independent of a
rescaling of the reddening measure \ESFD\ used to derive it, ie not affected by 
the 14\% downward rescaling of the \EBV\ measure of \cite{SchFin+98} that was 
suggested by \cite{SchFin11}.  Nonetheless the N(H)/\EBV\ ratio is an important 
quantity in its own right and it is not immune to uncertainty. 
\cite{NguDaw+18} recently derived 
N(H)/\EBV\ $= 9.4 \times 10^{21}$ mag\m1\ using 
2MASS-derived reddenings, equivalent to a 13\% downward rescaling of 
\ESFD.  \cite{LenHen+17} derived 
N(H)/\EBV\ $= 8.8 \times 10^{21}$ mag\m1\ from a larger-scale comparison, 
but {\it after} a 12\% downward rescaling, corresponding to the value
N(H)/\EBV\ $= 7.7 \times 10^{21}$ mag\m1\ of \cite{HenDra17}.

In Appendix A we showed that the most recent set of Arecibo singledish \HI\ 
emission-absorption measuremnents \citep{HeiTro03}, recently reviewed by 
\cite{FukHay+18}, gave systematically larger \WHI\ than those measured 
interferometrically, especially at smaller \WHI\ and \EBV. If this disparity could
be resolved, the sample of HI absorption measurements employed here might be
substantially expanded.

\acknowledgments

  The National Radio Astronomy Observatory is a facility of the National
  Science Foundation operated under contract by Associated  Universities, Inc.
  I thank the Hotel Bel \'Esperance in Geneva and the proprietor of Villa BEV 
  in Mont-Jean (St. Barthelemy) for their hospitality during the completion of 
  this manuscript.   The manuscript was improved by the anonymous referee's 
  comments and suggestions for improvement.



\appendix

\section{\WHI\ from emission-absorption experiments}

\cite{HeiTro03} carried on a long tradition of simultaneous  measurement
of \HI\ emission and absorption toward and around continuum sources at Arecibo 
\citep{DicTer+78,DicTer+79,PayTer+82,PaySal+83,ColSal+88}.  Their results, 
summarized and discussed recently by \cite{FukHay+18}, are shown here in 
Figure 4. The Arecibo emission-absorption experiment finds a flatter 
\WHI-\EBV\ relationship and consistently larger \WHI\ than the interferometer sample
at all \EBV. The regression line fit to the Arecibo data at \WHI\ $>$ 0.07 \kms\
is \WHI\ = $(17.0\pm1.6)$\ESFD$^{(0.975\pm0.049)}$ \kms, more nearly 
linear, but higher than that for the interferometrically-measured sample by factors of
2.19, 1.64 and 1.23 at \ESFD\ = 0.01, 0.1 and 1 mag, respectively. The
origin of this difference remains to be investigated as it has not previously
been recognized.

\cite{FukHay+18} tabulated values of the 353 GHz dust optical depth 
\taup\ for  the Arecibo emission-absorption sample in addition to \WHI\ 
so we generated \ESFD\ 
and compared \taup\ and \ESFD\ as shown in Figure 5.  With the exception of one
widely-discrepant sightline at \ESFD\ $\approx$ 2 mag, there is a tight
linear relationship.  The regression line fit
is \taup/$10^{-6} = (68.2\pm2.7)$\ESFD$^{(1.020\pm0.018)}$ and this
is only barely distiguishable from the \taup-\EQSO\ relationship
of \cite{PlaXI} using reddening values derived from photometry of
SDSS QSO.  That relationship, \taup $/10^{-6} = (67.1\pm1.3)$\EQSO\ 
is also shown in Figure 5.  The slope of a linear fit through the origin 
is  $<$\taup$>$/$<$\EBV$> = 67.1\times10^{-6}$ mag\m1.

What is surprising in Figure 5 is not the linear relationship, 
but that the \taup/\ESFD\ ratio is statistically indistinguishable from that 
predicted from the Planck \taup-\EQSO\ relationship of \cite{PlaXI}. 
\ESFD\ is equivalent to \EQSO\ for this eclectic collection of sightlines 
spanning quite a wide range of \EBV\ and a 14\% downward scaling of \ESFD\ is
manifestly inappropriate for this sample. 

\begin{figure}
\includegraphics[height=7cm]{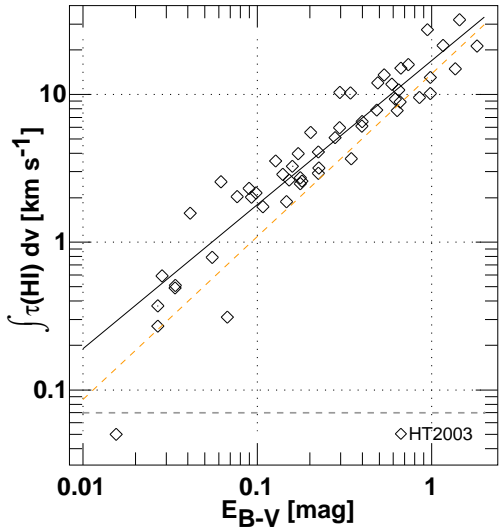}
 \caption[] {Integrated $\lambda$21 cm \HI\ optical depth from the Arecibo
emission-absorption survey of \cite{HeiTro03}.  The solid line is the
regression line using all data with \WHI\ $>$ 0.07 mag as in Figure 1,
\WHI\ = $(17.0\pm1.6)$\EBV$^{(0.975\pm0.048)}$ \kms. The 
dashed orange line is the power-law fit derived from the merged sample at 
right in Figure 1 with the same lower limit on \WHI.}
\end{figure}

\begin{figure}
\includegraphics[height=8.4cm]{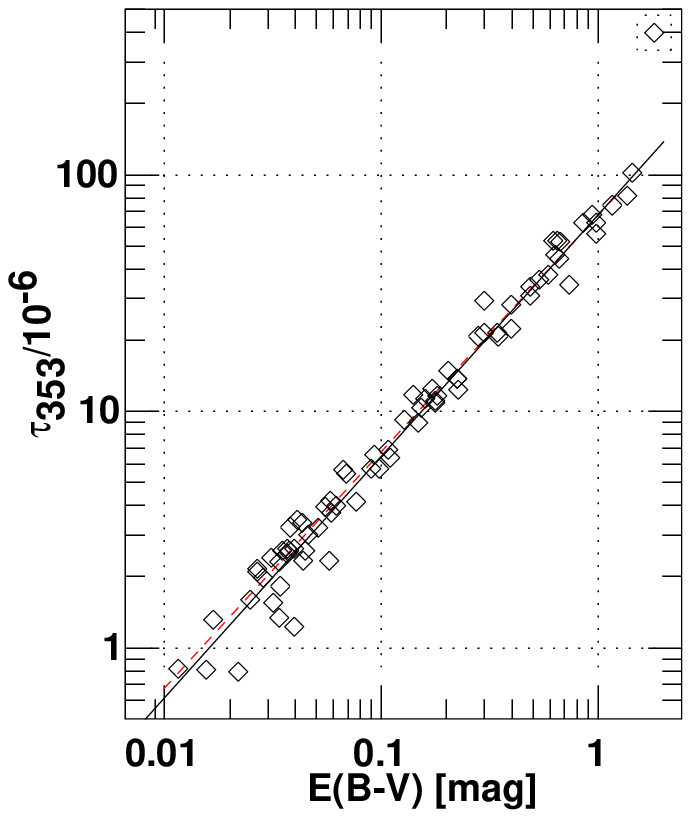}
 \caption[] {Planck 353 GHz dust optical depth \taup\ vs. \EBV\ from SFD98 
for 76 lines of sight observed in $\lambda$21cm \HI\ emission and absorption 
at Arecibo by \cite{HeiTro03} 
(see \cite{FukHay+18} for the values of \taup\ and \WHI).  
The regression relationship is 
$\taup/10^{-6} = (68.2\pm3.0) \EBV^{1.02\pm0.02}$ and the
barely-visible red dashed line is the mean relationship between \taup\ 
and reddening measured from quasar photometry by \cite{PlaXI}. The outlier
point at the highest \EBV\ was not used in the regression fit.}
\end{figure}


\bibliographystyle{apj}


\end{document}